# Scale-independent mixing laws

Bernard Montaron

**Abstract**

Mixing laws have been introduced in effective medium physics to calculate a bulk parameter of mixtures of several phases as a function of the parameter values and volume fractions for each phase. They have been successfully applied to derive mixture models for dielectric constant, thermal conductivity, electrical conductivity, etc.

Studied here are mixing laws that can be written in the form
$$f(\sigma) = a_1 f(\sigma_1) + a_2 f(\sigma_2) + ... + a_n f(\sigma_n)$$
with $a_1 + a_2 + ... + a_n = 1$ where $f$ is a continuous function applied to a bulk parameter $\sigma$ – e.g. the complex conductivity – and where the $a_i$ are the volume fractions of each phase. In the case of 'scale-independent' mixing laws i.e. mixing laws such that
$$\forall t > 0 \quad f(t\sigma) = a_1 f(t\sigma_1) + a_2 f(t\sigma_2) + ... + a_n f(t\sigma_n)$$
the function $f$ is shown to take only two forms: $f(x) = a \ln x + b$ or $f(x) = ax^p + b$. In other words scale-independent mixing laws can only be the geometric mean $\sigma = \sigma_1^{a_1} ... \sigma_n^{a_n}$ or the 'power mean' $\sigma = (a_1 \sigma_1^p + ... + a_n \sigma_n^p)^{1/p}$ with $p \neq 0$. The geometric mean corresponds to the limit case $p \to 0$. The other two limit cases are $\sigma = \max\{\sigma_1, \sigma_2, ..., \sigma_n\}$ for $p \to +\infty$ and $\sigma = \min\{\sigma_1, \sigma_2, ..., \sigma_n\}$ for $p \to -\infty$.

Consider the functional equation $\forall x, \forall y \in R \quad f(ax + by) = af(x) + bf(y)$ (1)
where $a$ and $b$ are positive real numbers such that $a + b = 1$ and where the unknown function $f$ is assumed to be continuous.

Put $g(x) = f(x) - f(0)$ then clearly $g$ is a continuous function and
$$\forall x, \forall y \in R \quad g(ax + by) = ag(x) + bg(y) \quad \text{with} \quad g(0) = 0 \quad (2)$$

**Lemma 1**

Any continuous function $g$ such that $g(0) = 0$ and
$$\forall x, \forall y \in R \text{ or } R^+ \quad g(ax + by) = ag(x) + bg(y)$$
for two fixed positive real numbers $a$ and $b$ is of the form $g(x) = cx$ for some real number $c$. Any continuous function $f$ such that
$$\forall x, \forall y \in R \text{ or } R^+ \quad f(ax + by) = af(x) + bf(y)$$
for two fixed positive real numbers $a$ and $b$ with $a + b = 1$ is of the form $f(x) = cx + d$ for some real numbers $c$ and $d$.

Note that in the first part of this lemma there is no need to assume $a + b = 1$.

Proof
Setting $y = 0$ in (2) leads to $\quad\quad \forall x \in R \quad\quad g(ax) = ag(x)$ (3)
and setting $x = 0$ in (2) leads to $\quad\quad \forall y \in R \quad\quad g(by) = bg(y)$ (4)
Using (3) and (4) in (2) leads to
$$\forall x, \forall y \in R \quad\quad g(ax + by) = g(ax) + g(by)$$ (5)
and finally putting $X = ax$ and $Y = by$ in (5)
$$\forall X, \forall Y \in R \quad\quad g(X + Y) = g(X) + g(Y)$$

This is Cauchy's equation for a continuous function. Therefore $g(x) = cx$ (see [1]) for some real number $c$. Setting $f(0) = d$ the general form for $f$ in (1) is $f(x) = cx + d$ and this completes the proof of lemma 1.

**Lemma 2**
Any continuous function $g$ such that $g(0) = 0$ and
$$\forall x_1, \forall x_2, ..., \forall x_n \in R \text{ or } R^+ \quad g(a_1 x_1 + a_2 x_2 + ... + a_n x_n) = a_1 g(x_1) + a_2 g(x_2) + ... + a_n g(x_n)$$
for fixed positive real numbers $a_1, a_2, ..., a_n$ is of the form $g(x) = cx$ for some real number $c$. Any continuous function $f$ such that
$$\forall x_1, \forall x_2, ..., \forall x_n \in R \text{ or } R^+ \quad f(a_1 x_1 + a_2 x_2 + ... + a_n x_n) = a_1 f(x_1) + a_2 f(x_2) + ... + a_n f(x_n)$$
for fixed positive real numbers $a_1, a_2, ..., a_n$ with $a_1 + a_2 + ... + a_n = 1$ is of the form $f(x) = bx + c$ for some real numbers $b$ and $c$.

Proof
Setting $x_3 = ... = x_n = 0$ (or more generally setting to zero $n - 2$ variables) and using lemma 1 leads directly to the proof of lemma 2.

Consider now the 'mixing law' applied to a positive physical parameter $\sigma$ (for example the electrical conductivity – see [2]) defined by
$$f(\sigma) = a_1 f(\sigma_1) + a_2 f(\sigma_2) + ... + a_n f(\sigma_n) \quad \text{with} \quad a_1 + a_2 + ... + a_n = 1$$ (7)

where $a_1, a_2, ..., a_n$ are the volume fractions of $n$ phases having physical parameter values $\sigma_1, \sigma_2, ..., \sigma_n$ respectively and where $\sigma$ is the physical parameter of the mixture medium obtained by mixing the phases according to a given 'mixing process'. The function $f$ is assumed to be continuous and in physics it will generally be monotonous.

Following [1] it is interesting to find out which functions $f$ give different values for $\sigma$ given $a_1, a_2, ..., a_n$ and $\sigma_1, \sigma_2, ..., \sigma_n$. To answer this question the following lemma will be used.

**Lemma 3**
The functional equation
$$\forall x_1,...,\forall x_n \in R \quad f^{-1}(a_1 f(x_1)+...+a_n f(x_n)) = g^{-1}(a_1 g(x_1)+...+a_n g(x_n)) \quad (6)$$
where $a_1, a_2,..., a_n$ are positive real numbers such that $a_1 + a_2 +...+ a_n = 1$ and where the unknown functions $f$ and $g$ are assumed to be continuous is equivalent to
$$\exists\, c \neq 0, \exists\, d \in R \quad \text{such that} \quad \forall x \in R \quad g(x) = cf(x) + d.$$

Proof
Start from $\forall x \in R \quad g(x) = cf(x) + d$. Setting $X = f(x)$
$x = f^{-1}(X) = g^{-1}(cX + d)$. Setting $X = a_1 f(x_1) +...+ a_n f(x_n)$
$$f^{-1}(a_1 f(x_1)+...+a_n f(x_n)) = g^{-1}(a_1 cf(x_1)+...+a_n cf(x_n) + d)$$
and since $d = a_1 d + a_2 d +...+ a_n d$ and $g(x_k) = cf(x_k) + d$ for all $k = 1, 2,..., n$
$$f^{-1}(a_1 f(x_1)+...+a_n f(x_n)) = g^{-1}(a_1 g(x_1)+...+a_n g(x_n))$$

Now conversely if
$$\forall x_1,...,\forall x_n \in R \quad f^{-1}(a_1 f(x_1)+...+a_n f(x_n)) = g^{-1}(a_1 g(x_1)+...+a_n g(x_n))$$
using $X_k = f(x_k)$ for all $k = 1, 2,..., n$. Equation (6) becomes
$$\forall X_1,...,\forall X_n \in R \quad F(a_1 X_1 +...+ a_n X_n) = a_1 F(X_1) +...+ a_n F(X_n)$$
where $F = g \circ f^{-1}$. According to lemma 2 there exist two real numbers $c$ and $d$ such that $F(X) = cX + d$. Setting $X = f(x)$ leads to $g(x) = cf(x) + d$ which completes the proof of lemma 3.

The generating functions in 'mixing laws' applicable to a given physical parameter – for a given phase mixing process – are all pair-wise linearly dependent. That property has profound implications as shown below.

In the mixing law (7) one should be able to change the physical units for the parameters $\sigma_1, \sigma_2,..., \sigma_n$ and this should result in the units for $\sigma$ to be changed accordingly. This would be automatically the case if the function $f$ included a reference parameter $\sigma_0$ such that $f(\sigma_k) = F(\sigma_k / \sigma_0)$ making the choice of units irrelevant. However in situations where there is no reason to distinguish any particular value $\sigma_0$ one must have the property
$$\forall t > 0 \quad f(\sigma) = a_1 f(\sigma_1) +...+ a_n f(\sigma_n) \Rightarrow f(t\sigma) = a_1 f(t\sigma_1) +...+ a_n f(t\sigma_n)$$

This is the case in particular for so-called 'scale-independent' problems. This property can be written in the following form
$$\forall t > 0 \quad tf^{-1}(a_1 f(\sigma_1) +...+ a_n f(\sigma_n)) = f^{-1}(a_1 f(t\sigma_1) +...+ a_n f(t\sigma_n))$$

**Lemma 4**
The only solutions of the functional equation
$$\forall t, \forall x_1, ..., \forall x_n \in R \quad t + f^{-1}(a_1 f(x_1) + ... + a_n f(x_n)) = f^{-1}(a_1 f(x_1 + t) + ... + a_n f(x_n + t))$$
where $a_1, a_2, ..., a_n$ are positive real numbers such that $a_1 + a_2 + ... + a_n = 1$ and where the unknown function $f$ is assumed to be continuous are
$$f(x) = ax + b \quad \text{or} \quad f(x) = as^x + b, \quad s > 0$$
where $a, b$ and $s > 0$ are real numbers.

Proof
Let $g(x) = f(x+t)$. Setting $y = f(x+t)$, $g^{-1}(y) = f^{-1}(y) - t$ and the functional equation in lemma 4 can be written in the form
$$\forall x_1, ..., \forall x_n \in R \quad f^{-1}(a_1 f(x_1) + ... + a_n f(x_n)) = g^{-1}(a_1 g(x_1) + ... + a_n g(x_n))$$

Using lemma 3 there exist real numbers $c \neq 0$ and $d$ independent of $x_1, x_2, ..., x_n$ such that $\forall x \in R \quad g(x) = cf(x) + d$. However $c$ and $d$ can be functions of $t$, therefore there exist functions $c(t)$ and $d(t)$ such that $\forall t, \forall x \in R \quad f(x+t) = c(t) f(x) + d(t)$ (7)

This is a special case of Vincze's equation [1] that is known to have only two sets of solutions $f(x) = ax + b$ or $f(x) = as^x + b$, $s > 0$ where $a, b$ and $s$ are real numbers.

**Theorem 1**
The only solutions of the functional equation
$$\forall t > 0, \forall x_1, ..., \forall x_n \in R^+ \quad tf^{-1}(a_1 f(x_1) + ... + a_n f(x_n)) = f^{-1}(a_1 f(tx_1) + ... + a_n f(tx_n))$$
where $a_1, a_2, ..., a_n$ are positive real numbers such that $a_1 + a_2 + ... + a_n = 1$ and where the unknown function $f$ is assumed to be continuous are $f(x) = a \ln x + b$ or $f(x) = ax^p + b$ where $a, b$ and $p$ are real numbers.

Proof
If one or several variables are equal to zero, the problem is the same with a lower value of $n$. It can therefore be assumed without loss of generality that all variables are positive i.e. non zero. Define the function $F$ such that $F = f \circ \exp$ and set $T = \ln t$, $X_k = \ln x_k$.
From there $F^{-1} = \ln \circ f^{-1}$ and the functional equation in the theorem becomes
$$T + F^{-1}(a_1 F(X_1) + ... + a_n F(X_n)) = F^{-1}(a_1 F(X_1 + T) + ... + a_n F(X_n + T))$$
for $\forall T, \forall X_1, ..., \forall X_n \in R$
Applying lemma 4 there are only two solutions for the function $F$:
$$F(X) = aX + b \quad \text{or} \quad F(X) = as^X + b, \quad s > 0$$
i.e. $\quad f(x) = a \ln x + b \quad \text{or} \quad f(x) = ax^p + b \quad \text{with} \quad p = \ln s$

Going back to the scale-independent mixing law $\sigma = f^{-1}(a_1 f(\sigma_1) + ... + a_n f(\sigma_n))$
satisfying $\forall t > 0 \quad tf^{-1}(a_1 f(\sigma_1) + ... + a_n f(\sigma_n)) = f^{-1}(a_1 f(t\sigma_1) + ... + a_n f(t\sigma_n))$
consider the two cases of this theorem:

First if $f(x) = a \ln x + b$ then setting $y = a \ln x + b$, $f^{-1}(y) = \exp((y-b)/a)$ and
$a_1 f(\sigma_1) + ... + a_n f(\sigma_n) = a(a_1 \ln \sigma_1 + ... + a_n \ln \sigma_n) + b$ therefore
$\sigma = \exp(a_1 \ln \sigma_1 + ... + a_n \ln \sigma_n) = \sigma_1^{a_1} \sigma_2^{a_2} ... \sigma_n^{a_n}$

Second if $f(x) = ax^p + b$ then $f^{-1}(y) = ((y-b)/a)^{1/p}$ and
$\sigma = (a_1 \sigma_1^p + a_2 \sigma_2^p + ... + a_n \sigma_n^p)^{1/p}$

As expected in both cases the parameters $a$ and $b$ disappear in the final expression.
This concludes the proof of theorem 2 below.

**Theorem 2**
A scale-independent mixing law i.e. a mixing law defined by the functional equation
$\forall t > 0 \quad f(\sigma) = a_1 f(\sigma_1) + ... + a_n f(\sigma_n) \Rightarrow f(t\sigma) = a_1 f(t\sigma_1) + ... + a_n f(t\sigma_n)$
with $a_1 + a_2 + ... + a_n = 1$ for some continuous function $f$ can only be :
    the geometric mean $\quad \sigma = \sigma_1^{a_1} \sigma_2^{a_2} ... \sigma_n^{a_n}$
    or the 'power mean' $\quad \sigma = (a_1 \sigma_1^p + a_2 \sigma_2^p + ... + a_n \sigma_n^p)^{1/p}$ with $p \neq 0$.

The geometric mean is the limit case of the power mean when $p$ tends to zero.
The arithmetic mean is obtained for $p = 1$, the harmonic mean for $p = -1$ and the
quadratic mean for $p = 2$, and that is the CRIM mixing law for complex conductivity
[2],[3],[4]. For $p \to +\infty$ and $p \to -\infty$ the limit cases of the power mean are
(respectively) $\max\{\sigma_1, \sigma_2, ..., \sigma_n\}$ and $\min\{\sigma_1, \sigma_2, ..., \sigma_n\}$ where 'max' and 'min' are
taken over all $\sigma_k$ with $a_k \neq 0$.

**Conclusion**

This mixing law is interesting in many respects for the modeling of the conductivity of
porous rocks. It contains Archie's equation as a special case (with equal Archie
exponents $n = m$) and it can be used to model petrophysical parameters of reservoir rocks
(see [5], [6]). It is important to mention that power mean mixing laws do not seem to
apply to electrical conductivity in the case of mixtures with at least one of the conductive
phase concentration below its percolation threshold. However it gives excellent results
for mixtures with fully connected phases.

# References

[1] Small, C.G., Functional Equations and How to Solve Them, Springer 2007, Problem Books in Mathematics

[2] Birchak, J., Gardner, L., Hipp, J., and Victor, J., 1974. High dielectric constant microwave probes for sensing soil moisture. Proceedings of the IEEE, vol.62 (1) pp.93-98

[3] Berryman, J.G., 1995. Mixture Theories for Rock Properties, Rock Physics and Phase Relations – A handbook of Physical Constants, pp.205 – 228

[4] Seleznev, N., Boyd, A., and Habashy, T., 2004. Dielectric Mixing Laws for Fully and Partially Saturated Carbonate Rocks, SPWLA 45th Annual Logging Symposium.

[5] Montaron, B., 2007. A Quantitative Model for the Effect of Wettability on the Conductivity of Porous Rocks. SPE 105041, MEOS March 2007.

[6] Montaron, B., 2009. Connectivity Theory – A New Approach to Model the Electrical Conductivity of Non-Archie Reservoir Rocks, *Petrophysics*, April 2009

---

Schlumberger, email: bmontaron@slb.com

Keywords: mixing laws, functional equations